\documentclass[twocolumn,prb,showpacs,preprintnumbers,amsmath,amssymb]{revtex4}
\usepackage{fancyhdr}
\usepackage{amssymb}
\usepackage{amsmath}
\usepackage[mathscr]{eucal}
\usepackage{latexsym}
\usepackage{amsbsy}
\usepackage{float}
\usepackage[dvips]{graphicx}
\usepackage{epsfig}
\usepackage{subfigure}
\usepackage{wrapfig}
\usepackage{bm}   
\usepackage{epsf}
\usepackage{float}

\newcommand{\be}{\begin{equation}}
\newcommand{\ee}{\end{equation}}
\newcommand{\bes}{\begin{equation*}}
\newcommand{\ees}{\end{equation*}}
\newcommand{\bea}{\begin{eqnarray}}
\newcommand{\eea}{\end{eqnarray}}
\newcommand{\bean}{\begin{eqnarray*}}
\newcommand{\eean}{\end{eqnarray*}}
\newcommand{\ba}{\begin{array}}
\newcommand{\ea}{\end{array}}

\newcommand{\Tr}{\mathrm{Tr}}

\begin{document}

\title{Magnetic superlattice and finite-energy  Dirac points in graphene}
\author{Luca Dell'Anna$^{1,2}$ and Alessandro De Martino$^{3}$}

\affiliation{
$^1$ Dipartimento di Fisica, Universit\`a di Trieste, I-34151, Italy\\
$^2$ Dipartimento di Fisica 'G. Galilei', Universit\`a di Padova, I-35131, Italy\\
$^3$ Institut f\"ur Theoretische Physik, Universit\"at zu K\"oln, 
D-50937 K\"oln, Germany}

\date{\today}

\begin{abstract}
We study the band structure of graphene's Dirac-Weyl quasi-particles 
in a one-dimensional magnetic superlattice formed by a periodic 
sequence of alternating magnetic barriers.
The spectrum and the nature of the states strongly depend
on the conserved longitudinal momentum and on the barrier width. 
At the center of the superlattice Brillouin zone
we find new Dirac points at finite energies where the dispersion 
is highly anisotropic, in contrast to the dispersion close to the neutrality point 
which remains isotropic. This finding suggests the
possibility of collimating Dirac-Weyl quasi-particles by tuning the doping.
\end{abstract}
\maketitle

\section{Introduction}
\label{sec1}

It is well-known that the low-energy electronic excitations in graphene
can be described as two flavors of Dirac-Weyl (DW) quasi-particles,
whose linear spectrum and chiral nature underly many of the unsual and intriguing
properties of this new material.\cite{reviews}
The prospect of employing graphene as a building block in electronic 
nanodevices has stimulated an intense research activity addressing the 
problem of how to manipulate its peculiar electronic band structure.  
A great deal of attention has been recently devoted 
to {\em superlattice structures}, where external spatially periodic electric 
or magnetic fields are applied to a graphene monolayer. 
In many cases the potential modulations
are smooth and their spatial period greatly exceeds the lattice costant, 
so that the quasi-particle dynamics is well described by 
an effective DW Hamiltonian in the presence of external fields.
In the case of electric superlattices interesting new features have been theoretically
predicted, as the phenomenon of supercollimation\cite{park2008b,park2008c} 
and the emergence of new zero-modes,\cite{park2008a,park2009,brey2009,barbier2010}
i.e., additional zero-energy DW quasi-particles induced in the vicinity of the superlattice 
Brillouin zone (SBZ) boundary. 

In this paper we focus on the electronic properties of one-dimensional 
magnetic superlattices (1D MSL). There exists to date, to the best of our knowledge, 
no experimental realization of such structures. However, 
there is no principle obstruction to the fabrication of  magnetic 
potentials in graphene that vary on submicrometer scales, by using techniques well 
established in the case of the two-dimensional electron gas in semiconductor 
heterostructures.\cite{nogaret2010} 
Moreover, local strain in graphene induces a spatially varying 
pseudo-magnetic field, and recent experimental results\cite{bao2009} 
indicate that one can achieve a rather high degree of control over the strain.
For example, it is possible to produce and control a periodic pattern of ripples,\cite{bao2009}  
which opens an alternative way to the realization of a MSL by 
strain engineering.\cite{guinea2008,pereira2009}
We thus expect that graphene MSL will be available in the near future.
 
There already exists a number of theoretical works which have investigated some properties
of MSL. In Ref.~\onlinecite{luca2009} we found that, quite surprisingly, in a 1D MSL
the Fermi velocity at the Dirac points is isotropically renormalized, in strong contrast to 
the case of 1D electrostatic superlattices, where the renormalization
is strongly anisotropic.\cite{park2008b,park2008c}
The same result was independently found by Snyman, \cite{snyman2009} who focused on 
the general question, under which conditions a spectral gap opens in the presence of 
periodic magnetic and electric fields, and by Tan et al., \cite{louie2010} which showed
that  the problem of a 1D MSL can be mapped to that of an electric superlattice. 
Other works\cite{ghosh2009,ramezani2009,ramezani2010} studied the special 
case of a magnetic  Kronig-Penney potential with delta-function barriers,
emphasizing the analogies to the optical properties of a medium with a
periodic modulation of the refractive index. 
The generation of new zero-energy Dirac points in a staggered magnetic field 
and the implications of the snake states on the integer quantum Hall effect in graphene
have been discussed in Refs.~\onlinecite{xu2010a} and \onlinecite{xu2010b}.
Recently, the phase-coherent transport in a strain-induced periodic 
pseudo-magnetic field has also been studied.\cite{belzig2010}

Here we discuss in detail a complementary aspect, which apparently
has not been noticed so far, namely, the existence of additional 
finite-energy Dirac points in the spectrum of a 1D MSL
at the center of the 1D superlattice Brillouin zone. 
We shall see that in the vicinity of these new points the dispersion 
has a highly anisotropic double-cone shape, indicating the possibility of 
achieving a high degree of collimation by tuning the doping. 

The rest of the paper is organized as follows. In Sec. \ref{sec2} we present
the model and formulate the basic equation for the exact calculation of  the 
band structure. The spectrum close to zero energy is briefly reviewed in 
Sec. \ref{sec3}, while in Sec. \ref{sec4} we discuss the general 
numerical solution of the spectral equation and the new Dirac points emerging 
at finite energies. In Sec. \ref{sec5} we provide an explicit analytic solution of the spectral 
equation in two asymptotic regimes. Sec. \ref{sec6} is devoted to the perturbative calculation 
of the spectrum in two limiting cases, which gives additional physical insights 
into the nature of the superlattice quantum states. Finally, Sec. \ref{sec7} 
presents some conclusions.

\section{The model}
\label{sec2}

We consider a magnetic field configuration uniform in the $y$-direction and staggered 
in the $x$-direction on a  length scale much larger than the lattice constant.  
The smoothness of the vector potential allows us to neglect intervalley 
scattering and to use the single-valley continuum DW theory. 
At the same time, at low energies the typical  
de Broglie wavelength of 	quasi-particles is much larger than the length 
scale over which the magnetic field varies, and we can approximate the 
magnetic profile as piecewise constant. Since the Zeeman effect is
very small in graphene we shall neglect all spin effects. 
Then including the perpendicular magnetic field via minimal coupling, the 
DW equation reads 
\begin{equation}
v_F {\bm \sigma} \cdot \left(-i{\bf \hbar \nabla} + \frac{e}{c} 
{\bf A}\right) \Psi = E \Psi ,
\label{dirac2d}
\end{equation}
where ${\bm \sigma}= (\sigma_x,\sigma_y)$ are Pauli matrices acting
in sublattice space, and $v_F=8\times10^5\,$m/s is the 
Fermi velocity. 
In the Landau gauge, 
${\bf A}=(0, A(x))$, with $B_z = \partial_x A$, the $y$-component of the
momentum is a constant of motion, and the spinor
wavefunction can be written as $\Psi(x,y) = \psi(x)e^{ik_yy}$, 
whereby Eq.~(\ref{dirac2d})  is reduced to a one-dimensional problem:
\begin{align}
H \psi &= E \psi,  \label{dirac1d} \\
H &= -i \left( 
\begin{array}{cc}
0 & \partial_x + k_y + A(x)  \\
\partial_x  - k_y - A(x) & 0
\end{array}
\right) .
\label{ham1d}
\end{align}
Equations (\ref{dirac1d}) and (\ref{ham1d}) are 
written in dimensionless units: with $B>0$  denoting 
the typical magnitude of the magnetic field and 
$\ell_B=\sqrt{\hbar c/e B}$ the associated
magnetic length, we express the vector
potential $A(x)$ in units of $B\ell_B$, the energy $E$ 
in units of $\hbar v_F/\ell_B$, and $x$ and $k_y$ respectively 
in units of $\ell_B$ and  $\ell^{-1}_B$.  
The values of local magnetic fields 
in the barrier structures produced by
ferromagnetic stripes range up to 1 T, 
with typical values of the order of tenth of Tesla. Thus typical length and 
energy scales in this problem are given, for $B\approx 0.1\,$T, by
$\ell_B \approx 80\,$nm and $\hbar v_F/\ell_B \approx 7\,$meV.

We shall consider a periodic magnetic profile whose elementary 
unit is given by a magnetic barrier ($B_z=B$) of width $d$
followed by a magnetic  well ($B_z=-B$) of the same width.\cite{luca2009} 
Thus the net magnetic flux through the unit cell vanishes. 
The vector potential is accordingly chosen as 
\begin{equation}
A(x) = \left\{
\begin{array}{ll}
x-x_{j}-\frac{d}{2}, &  \;\; x\in [x_{j}, x_{j}+ d], \\
\frac{3d}{2}+x_{j}-x, &   \;\; x\in [x_{j}+d, x_{j+1}],
\end{array}
\right.
\label{gauge2}
\end{equation}
where $j\in \mathbb{Z}$ and $x_{j}=2dj$. 
After solving the DW equation in the presence of a constant magnetic field,\cite{ale}
it is convenient to define two matrices whose columns are given by the 
(unnormalized) eigenspinors
in the regions of positive and negative magnetic field:
\begin{equation}
\label{WB}
{\cal W}_B(x) =
\left(
\begin{array}{ll}
D_p(q) & D_p(-q)\\
\frac{i\sqrt{2}}{E}D_{p+1}(q) &\frac{-i \sqrt{2}}{E}D_{p+1}(-q)
\end{array}
\right)
\end{equation}
for $ x\in [x_{j}, x_{j}+ d]$ 
and
\begin{equation}
\label{W-B}
{\cal W}_{-B}(x) =
\left(
\begin{array}{ll}
D_{p+1}(-q) & D_{p+1}(q)\\
\frac{-iE}{\sqrt{2}}D_{p}(-q) &\frac{iE}{\sqrt{2}}D_{p}(q)
\end{array} \right)
\end{equation}
for  $x\in [x_{j}+d, x_{j+1}]$,
where we use the notation $q=\sqrt{2}(A(x)+k_y)$, $p=E^2/2-1$, 
and $D_p(q)$ is the parabolic cylinder function.\cite{grad}
According to Eq.~(\ref{gauge2}) we have $A(0)=-d/2$ and $A(d)=d/2$. 
Imposing periodic boundary conditions on the wavefunction 
implies a quantization condition for the energy, which is 
found to be\cite{luca2009}
\be
\label{sc}
2\cos(2 d k_x)=\Tr \, \Omega(E,k_y,d),
\ee 
where $k_x$ is the 1D quasimomentum ranging in the SBZ, 
$-\frac{\pi}{2d} < k_x \leq \frac{\pi}{2d}$,
and the matrix $\Omega$
reads
\be
\label{Omega}
\Omega(E,k_y,d)={\cal W}_B^{-1}(0){\cal W}_{-B}(0)
{\cal W}_{-B}^{-1}(d){\cal W}_B(d).
\ee
Formula (\ref{sc}) is the basic equation which determines the MSL band structure.
Its solutions will be discussed in detail in the rest of the paper. 

Before closing this section, 
we notice that the energy spectrum $E(k_x,k_y)$
is obviously an even function of $k_x$
and is also an even function of $k_y$. 
This follows from the fact that, since $A(x)=-A(d-x)$, if 
$\psi_{k_y}(x)e^{ik_yy}$ is a solution of the  DW equation of energy
$E$ then $\sigma_z \psi_{-k_y}(d-x)e^{-ik_yy}$ is a solution with the same 
energy. This symmetry implies that the states at $k_x=k_y=0$
are doubly degenerate and underlies the existence of the finite-energy 
Dirac points. Moreover the  particle-hole symmetry of the DW equation 
implies that the band structure is symmetric under reflection about $E=0$ 
and therefore we will mostly focus on the non-negative part of the spectrum. 

\section{Neutrality point and group velocity}
\label{sec3}

\begin{figure}[h!]
 \includegraphics[width=0.44\textwidth]{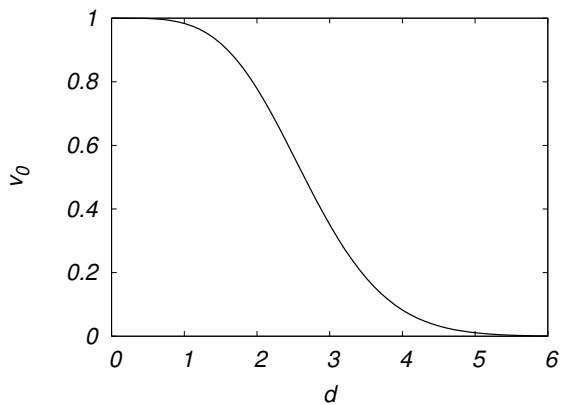}
\caption{The group velocity $v_0$ (in units of $v_F$)
at the neutrality point as a function of $d$ (in units of $\ell_B$).}
\label{fig.1}
\end{figure}

To begin with, we briefly consider the structure of the dispersion in the vicinity of
the neutrality point, i.e., close to zero energy.
Surprisingly enough, despite the strong anisotropy of the magnetic profile, 
the dispersion presents a Dirac cone with an isotropically renormalized velocity.
\cite{luca2009,snyman2009,louie2010}
To see this, we notice that $ \Tr \, \Omega (E=0,k_y,d) =2 \cosh(2d k_y) $, which can be easily checked by the explicit calculation of the zero-energy states,
and by further expanding  the trace to lowest order in $E$ and $k_y$ we obtain
\be
\Tr \, \Omega(E,k_y,d) \simeq 2+4 d^2k_y^2 -{\cal K}_0(d)\, E^2 .
\ee
The coefficient of the $E^2$ term is given by
\begin{align}
{\cal K}_0(d)  = \frac{1}{\pi e^{d^2/4}}&
\left[ \pi e^{3d^2/8}
\text{erf} \left(d/2\right) - D^{(1,0)}_0\left(-\frac{d}{\sqrt{2}}\right) + \right. \nonumber\\
& \left. + D^{(1,0)}_0\left(\frac{d}{\sqrt{2}}\right) \right]^2,
\end{align}
where ${\text{erf}}\left( x \right)$ 
is the error function\cite{grad} and $D^{(1,0)}_p(z)\equiv \partial_p D_p(x)$ denotes
the derivative of $D_p(x)$ with respect to the index.
Expanding also the right-hand side of Eq. (\ref{sc})  to lowest order in $k_x$
we finally get the dispersion 
\be
\label{disp}
E(k_x,k_y) = \pm \, v_0(d) \sqrt{k_x^2 + k_y^2},
\ee
with the $d$-dependent group velocity $v_0(d)$ given by
\be
\label{v0}
v_0=\frac{2d}{\sqrt{{\cal K}_0(d)}}.
\ee
The group velocity is plotted in Fig.~\ref{fig.1}, which shows that
$v_0(d)$ is always smaller than the Fermi velocity ($v_F=1$ in our units).
It monotonously decreases for increasing $d$, which can be easily understood, 
as the states become more and more localized inside the magnetic regions
(see Sec. \ref{sec6}), and for $d\gg 1$ we find
\begin{align}
v_0(d)\simeq \frac{2d}{\sqrt{\pi}} e^{-d^2/4}.
\label{vlarged}
\end{align}
For $d\ll 1$ we find instead 
\begin{align}
v_0(d)\simeq 1-d^4/60.
\label{vsmalld}
\end{align}
Thus, restoring the units, $d \rightarrow d/\ell_B \propto d \sqrt{B}$, we see that 
the correction to the Fermi velocity for small magnetic field is quadratic in $B$.

\section{Dirac points at finite energies}
\label{sec4}

\begin{figure}[h!]
\includegraphics[width=0.44\textwidth]{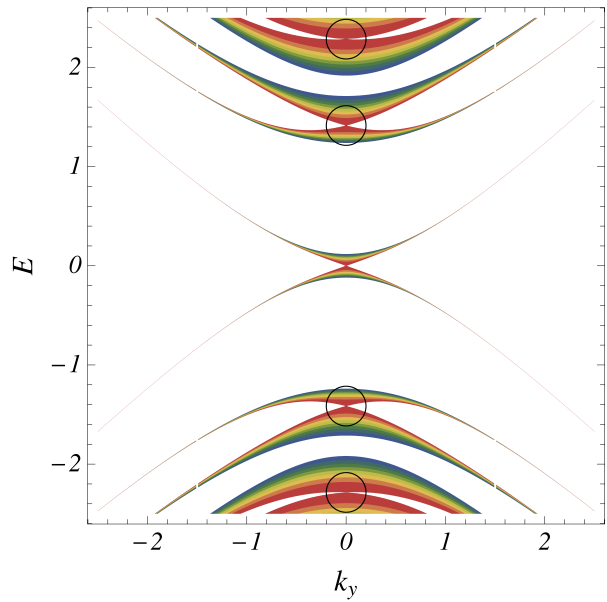}
\caption{(Color online) Contour plot of $\Tr\,\Omega(E,k_y,d)$ as function of 
$E$ and $k_y$ for $d=3$, with values in the range $[-2,2]$, increasing from blue to red. 
The circles emphasize the finite-energy Dirac points.}
\label{fig.2}
\end{figure}
Let us  now discuss the full band structure. Figure~\ref{fig.2} presents
a contour plot of $\Tr \, \Omega$, where the values outside the physical range
$[-2,2]$ are excluded. One recognizes electronic bands that narrow upon increasing
$|k_y|$. Physically, this corresponds to the crossover from states at small $k_y$,
predominantly localized inside the magnetic regions (broadened Landau levels),
to states at large $|k_y|$, localized at the interfaces where the magnetic 
field changes sign, the so-called "snake states".\cite{tarun,cserti}  
Qualitatively, this picture can be easily understood by looking 
at the profile of the effective potential in the Schr\"odinger equation 
satisfied by the two components of the DW spinor,
$V_{\text{eff}}(x) = \sigma_z B_z(x) + \left[ A(x)+k_y\right]^2$.
For $|k_y| \ll d$ the effective potential 
presents a periodic sequence of approximately parabolic wells whose bottoms 
are alternately shifted by $\pm B$ and, for $d\gg 1$, are located deep inside 
large magnetic regions. The corresponding eigenstates
are thus close to Landau states.  For large $|k_y|$, instead,
the potential has deep minima at $x=2n d$ for $k_y>0$  and $x=(2n+1)d$  for $k_y<0$, 
and localizes the states respectively at the interfaces $-B/+B$ and $+B/-B$, resulting in
 snake states propagating in the positive and negative $y$-direction.

\begin{figure}[t!]
 \includegraphics[width=0.44\textwidth]{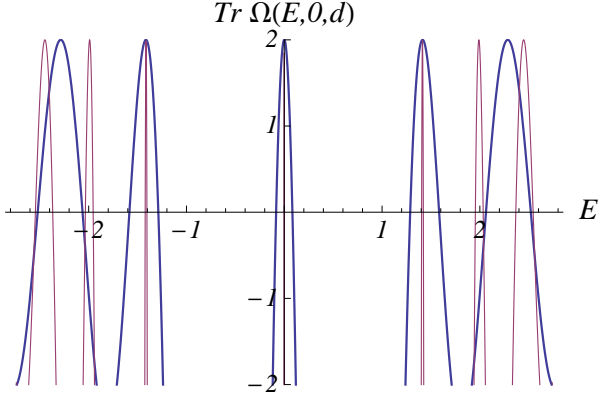}
\caption{(Color online) Plot of $ \Tr \, \Omega(E,k_y,d)$ as function of $E$ at $k_y=0$ 
and $d=3$ (blue thick line) and $d=5$ (magenta thin line), plotted within the 
physical range $[-2,2]$.}
\label{fig.3}
\end{figure}

\begin{figure}[t!]
\includegraphics[width=0.44\textwidth]{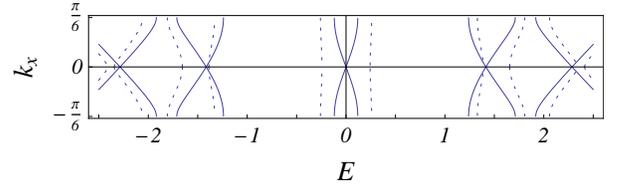}
\caption{Plot of $k_x$ versus $E$ for $d=3$, at $k_y=0$ (solid line) and 
$k_y=0.6$ (dashed line).}
\label{fig.4}
\end{figure}

Inspection of Fig.~\ref{fig.2} shows that at $k_y=0$ 
finite-energy degeneracy points exist, where a DW-like structure, i.e., 
a double-cone dispersion, seems to appear. We then focus on the region
close to $k_y=0$.
The plot of $\Tr \, \Omega(E,0,d)$ as a function of $E$ (see Fig.~\ref{fig.3})
indicates that for any $k_x$ in the 1D SBZ  there are infinite 
pairs of solutions $(E^+_n(k_x),E^-_n(k_x))$, $n\in \mathbb{Z}$. At the zone center
$k_x=0$ the solutions coincide pairwise, $E^+_n(0)=E^-_n(0)$, and the corresponding 
states are doubly degenerate. The degeneracy is lifted by a finite value of $k_y$
(see Fig.~\ref{fig.4}). Moreover in the limit of large $d$ the difference
$E^+_n(k_x)-E^-_n(k_x)$ tends to zero and 
the energy eigenvalues converge toward the Landau level values 
$E^\pm_n(0)\rightarrow \text{sign}(n)\sqrt{2|n|}$.
These qualitative considerations can be made precise by the exact 
numerical solution of Eq.~(\ref{sc}) (see below) and by the perturbative 
analysis of the spectrum (see Sec. \ref{sec6}).

\begin{figure}[h!]
\includegraphics[width=0.44\textwidth]{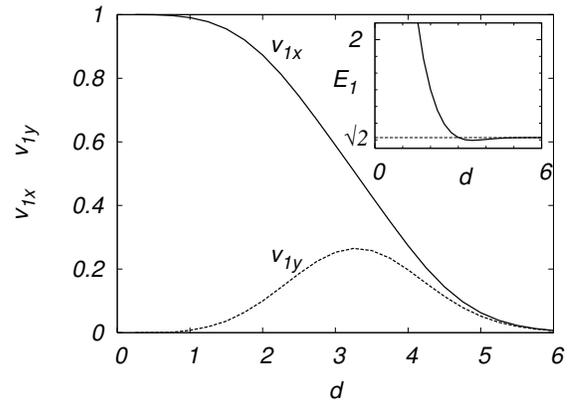}
\caption{Plot of the velocities $v_{1 x}$ and $v_{1 y}$  (in units of $v_F$)
as functions of $d$ (in units of $\ell_B$), at the first finite-energy 
Dirac point $E_1$. In the inset the plot of $E_1$ (in units of $\hbar v_F/\ell_B$) 
as a function of $d$ (in units of $\ell_B$).}
\label{fig.5}
\end{figure}

At $k_y=0$ the trace in Eq.~(\ref{sc}) can be rewritten as 
\be
\Tr \, \Omega(E,0,d)= 2-R^2(E,d),
\ee 
where $R(E,d)$ is a real function defined as
\be
R(E,d)=\frac{ \sum_{r=\pm1} r \left[ D^2_{p+1}(-r d/\sqrt{2}) +
(1+p)  D^2_{p}(r d/\sqrt{2})\right]}{\sqrt{1+p}
\sum_{r=\pm 1}
D_{p+1}(rd/\sqrt{2})D_{p}(-rd/\sqrt{2})}.
\ee
The quantization condition (\ref{sc}) at $k_x=k_y=0$
thus reduces to 
\be
R(E,d)=0,
\label{qc}
\ee
which can be easily solved numerically.
Due to the particle-hole symmetry of the DW equation (\ref{dirac2d}), 
the solutions of Eq.~(\ref{qc}) always occur in pairs $\pm E_n$, $n=0,1,2\dots$.
By expanding the trace around any $E_n$ we find at leading order
\be
\Tr \, \Omega(E,k_y,d) \simeq 2- {\cal K}_n\, (E-E_n)^2+c_n\, k_y^2,
\label{tracen}
\ee
where we define
\begin{align}
{\cal K}_n &=-\frac{1}{2}\frac{\partial^2}{\partial E^2} 
\Tr \, \Omega(E,0,d))\Big|_{E=E_n}, \label{kappan}\\
c_n & =\frac{1}{2}\frac{\partial^2}{\partial k_y^2} \Tr\,\Omega(E_n,k_y,d)\Big|_{k_y=0}.
\label{cn}
\end{align}
Therefore, from Eq.~(\ref{sc}) we obtain, in analogy to Eq.~(\ref{disp}), 
 the anisotropic Dirac-like dispersion 
\be
\label{dispersion}
E(k_x,k_y)=E_n \pm \sqrt{v_{n x}^2\, k_x^2+v_{n y}^2\,k_y^2},
\ee
with 
\begin{align}
v_{n x}=\frac{2d}{\sqrt{{\cal K}_n}}, \quad\quad
v_{n y}=\sqrt{\frac{c_n}{{\cal K}_n}}.
\end{align}
For instance, for $d=3$ the first non-vanishing solution 
of Eq.~(\ref{qc}) is $E_1\approx  1.4145269$, for which 
${\cal K}_1\approx 103.65$ and $c_1\approx 6.63$. Consequently the 
velocities are $v_{1 x} \approx 0.59$ and 
$v_{1 y} \approx 0.25$. 
The second solution is $E_2\approx 2.2854943$ for which 
$v_{2 x}\approx 0.88$ and $v_{2 y}\approx 0.06$, and so on. 
Focussing on the first Dirac point above the zero-energy one, 
the group velocities in the $x$ and $y$ directions are plotted in Fig.~\ref{fig.5}
as function of $d$. We notice that there exists a range of values where
$v_{1x}$ is only weakly renormalized, whereas $v_{1y}$ is strongly suppressed.
The same occurs also at the higher Dirac points. This quite unexpected result implies 
that the 1D MSL hinders the propagation of the quasi-particles in the direction normal 
to it and thus produces a certain degree of collimation. 

Before closing this section, we observe that
the Taylor expansion of $c_n$ for small $d$ reads 
$c_n=4d^2-\frac{8}{3}E_n^2d^4+\frac{8}{15}E_n^4d^6-\frac{1}{315}E_n^2(11+16E_n^4)d^8+..$.
However we will see in the following that the dimensionless energies for $d\rightarrow 0$
diverge as $E_n\sim 1/d$. Therefore all the terms in the expansion are of the same order, 
which suggests that a perturbative calculation of $v_{ny}$ could be problematic. This is indeed
the case, as we will see in Sec. \ref{smallB}.

\section{Asymptotic behaviors}
 \label{sec5}
 
 In this section we complement the previous discussion by the explicit 
analytic solution of Eq. (\ref{sc}) in two limiting cases, namely, 
{\em i)} at large energy and {\em ii)} when the barrier width is much larger 
than the magnetic length.

\subsection{High energies or vanishing magnetic field}
  
For large values of the energy $E$ we can simplify the expression of $\Tr \, \Omega$ 
by using the asymptotic behavior of the parabolic cylinder function
for large values of the index $p$:\cite{watson,lucaspin} 
\begin{align}
D_p(z) &\simeq \sqrt{2} \cos \left( \frac{\pi p}{2}- z\sqrt{p}\right) (p/e)^{p/2}, \label{asy1} \\
D_{p+1}(z) &\simeq - \sqrt{2} \sqrt{p} \sin \left( \frac{\pi p}{2}- z\sqrt{p}\right) (p/e)^{p/2}, 
\label{asy2}
\end{align}
and we obtain the simple expression
\be
\label{Tr_apx_1}
\Tr \, \Omega(E,k_y,d)\,\simeq\, 2\cos(2d E).
\ee
The solutions of Eq.~(\ref{qc}) are then given by
\be
E_n \simeq \pm \frac{\pi n}{d}.
\label{Bzero}
\ee
These values are easily understood for vanishing magnetic field. 
In that case, in fact, Eqs. (\ref{asy1}) and (\ref{asy2}) remain valid 
(provided $k_y\rightarrow 0$), since in our units $E\propto 1/\sqrt{B}$.
The energies in Eq. (\ref{Bzero}) then are nothing but the crossing points 
at $k_x=0$ of the unperturbed conical dispersion 
folded along the $k_x$-direction into the SBZ. 
Using Eqs.~(\ref{kappan}) and (\ref{cn}), at the energies $E_n$
we get the following limiting values for $v_{n x}$ and $v_{n y}$:
\be
\label{vx_asy}
v_{n x}\rightarrow 1,\quad\quad v_{n y}\rightarrow 0,
\ee 
for $n> 0$. This result has to be contrasted with the case $n=0$ 
(at the neutrality point) where $v_{0 x}=v_{0 y}\rightarrow 1$, 
as shown in Sec.~\ref{sec3}. Indeed, for large energies 
the $k_y$-dispersion around $E_{n>0}$ flattens, as one can see from the
fact that Eq.~(\ref{Tr_apx_1}) does not depend on  $k_y$. Therefore 
the asymptotic behavior for large $n$ is $v_{n y}\rightarrow 0$. 
Eq.~(\ref{vx_asy}) is confirmed by
the exact results obtained by keeping $d$ fixed and increasing $n$. 
For example, at $d=3$ and for $n=0,\, 1,\, 2$ 
we find $v_{n x}\approx 0.35,\, 0.59,\, 0.88$ and $v_{n y}\approx 0.35,\, 0.25,\, 0.06$.

\subsection{Large magnetic field or large $d$}
\label{larged} 

In the limit of very large barrier width, or equivalently of very large 
magnetic field, $d\gg 1$,  we expect that the spectrum reduces to 
doubly degenerate Landau levels.
To see this, we notice that since $d$ appears in the argument of the 
parabolic cylinder functions, we need their asymptotic behavior for large 
values of the argument. For  $z$ real and positive one has the 
following asymptotic expressions\cite{grad}
\bea
&&D_p(z)\simeq e^{-z^2/4} z^p,\\
&&D_p(-z)\simeq e^{-z^2/4} (-z)^p-\frac{\sqrt{2\pi}e^{i\pi p}e^{z^2/4}}{(-z)^{p+1}\Gamma[-p]},
\eea
where $\Gamma$ is the Gamma function.
In this case $\Tr \, \Omega(E,0,d)$ reduces to
\be
\Tr \, \Omega(E,0,d) \,\simeq \,2\cos(\pi E^2)-
\frac{\pi \,2^{E^2} E^2e^{d^2/2}}{d^{2E^2}\Gamma^2[1-E^2/2]}, 
\label{asytrace}
\ee
and the solutions of Eq.~(\ref{qc}) are just the Landau levels
\be
E_n\simeq \pm\sqrt{2 n}, \quad n=0,1,2,\dots
\ee
From Eq.~(\ref{asytrace}) and Eq.~(\ref{kappan}) 
we calculate
\be
{\cal K}_n \simeq \left\{
\ba{ll}
\pi e^{d^2/2},\;\; & {\textrm {for}} \; n=0,\\
4\pi\,\frac{4^{n}}{d^{4n}} (n!)^2 e^{d^2/2}, \; & {\textrm {for}} \; n>0,
\ea\right. 
\ee
while from the asymptotic form of $\Tr \, \Omega(E_n,k_y,d)$ 
(whose lenghty expression is not reported here) and Eq.~(\ref{cn})
we find
\be
c_n \simeq \,4 d^{2}(1-4n/d^2)^2.
\ee
Consequently, we obtain, for $n=0$, $v_{0 x}=v_{0 y}\simeq \frac{2d}{\sqrt{\pi}}e^{-d^2/4}$, 
in agreement with Eq.~(\ref{vlarged}), and for $n>0$ the following asymptotic velocities:
\bea
&&v_{n x} \simeq 
\frac{d^{2n+1}e^{-d^2/4}}{2^n n!\sqrt{\pi }},
\label{v_nx_asy}\\
&&v_{n y} \simeq v_{n x} |1-4n/d^2|.\phantom{\Big |}
\label{v_ny_asy}
\eea
Notice that as $d\rightarrow \infty$ we get 
$v_{n y}\rightarrow v_{n x}$, namely, the velocities vanish 
exponentially while recovering the isotropy, as one can see
in Fig.~\ref{fig.5}.


\section{Perturbative approach}
\label{sec6}

In this section we complement the results obtained
above by the explicit analytic computation of the spectrum 
in two limiting cases, where a perturbative approach can be used, 
The perturbative parameter is the ratio $d/\ell_B$ between barrier 
width and magnetic length.\cite{footnote} In our units the parameter is simply $d$ 
and the two perturbative regimes are respectively $d\ll 1$ and $d\gg 1$. 
In Sec. \ref{smallB} we treat the case of small magnetic field and/or small width, $d\ll 1$,
where the magnetic modulation is a weak periodic perturbation imposed on 
freely propagating DW quasi-particles.
In Sec. \ref{largeB} we consider, instead, the case of large magnetic field and/or  
large barrier width, $d \gg 1$. In this "atomic" limit, the unperturbed states 
are two sets of degenerate (relativistic) Landau orbitals 
localized respectively in the center of the regions of positive and negative magnetic field,
and the perturbation is the hopping between nearest-neighbor orbitals. 

We shall see below that the existence of finite-energy Dirac points is 
not captured by the lowest-order perturbative calculation in the weak periodic 
modulation regime, but it is nicely confirmed by the tight-binding-like analysis 
in the opposite regime of large $d$.

\subsection{Case $d\ll1$}
\label{smallB}

Following standard steps\cite{ashcroft} we make the ansatz
\be
\psi(x) = \sum_{\kappa_x} c_{\kappa_x} e^{i\kappa_x x},
\ee
where $\kappa_x=k_x-K_n$, with $k_x$ in the SBZ, 
$-\frac{\pi}{2d} < k_x \leq  \frac{\pi }{2d}$,
and $K_n$ a reciprocal lattice vector, $K_n=\frac{\pi n}{d}$, $n\in \mathbb{Z}$.
The DW equation is then equivalent to 
\begin{align}
\left[ \sigma_x (k_x-K_n) + \sigma_y k_y  - E \right]   & c_{k_x-K_n} + \nonumber \\
+ \sum_{m\in \mathbb{Z}} A_{K_m-K_n} & \sigma_y c_{k_x-K_m}=0,
\label{eq1}
\end{align}
where the Fourier components of $A(x)$ are given by
\begin{align}
A_{Q}  & = \frac{1}{2d} \int_{-d}^d dx \,e^{-iQx} A(x)=  \nonumber \\
 & =\frac{ 2(1-\cos Qd)-dQ\sin Q d}{2dQ^2} = \nonumber \\
& =\left\{
\begin{array}{ll}
0 & \text{for} \; Q=\frac{2n \pi}{d} ,\\
\frac{2}{dQ^2} & \text{for} \; Q=\frac{(2n+1) \pi}{d} .
\end{array}
\right.
\end{align}
Thus the periodic potential in Eq. (\ref{eq1}) couples a 
state with $K_n$ even to all states with $K_m$ odd and viceversa,
but the coupling $A_{K_m-K_n}$ rapidly decreases for increasing 
momentum transfer as $1/(K_m-K_n)^2$. 

We now focus on $ k_x \simeq 0$, i.e.,  
on the spectrum close to the center of the SBZ. 
The pairwise quasi-degenerate states at $K_n$ and $K_{-n}$
are never mixed by the potential since $A_{K_{-n}-K_{n}}=A_{-\frac{2n \pi }{d}}=0$. 
All other states are non-degenerate.
Hence the leading energy correction for a state of unperturbed energy 
$E^0_n(k_x,k_y) = \sqrt{(k_x-K_n)^2 + k_y^2}$ 
 is of second-order and reads
\begin{align}
\delta E_n(k_x,k_y) = \sum_{m\in \mathbb{Z},r =
\pm} \frac{\left| \langle K_n,+ \right| A_{K_m-K_n} \sigma_y
\left|  K_m, r \rangle
\right|^2}{E_n^0- r E^0_{m}}.
\end{align}
With the state $|K_n,r\rangle$ given by
\be
|K_n,r\rangle =\frac{1}{\sqrt{2}} \left(
\begin{array}{c}
1\\
r\frac{(k_x-K_n)+ik_y}{E^0_n}
\end{array}
\right),
\ee
after simple algebra 
we obtain the compact expression
\begin{align}
\delta E_n(k_x,k_y) = d^4 {\cal R}(d|k_x-K_n|) E_n^0(k_x,k_y),
 \label{pertsmalld}
\end{align}
where we have introduced the function
\be
{\cal R}(z)= \frac{1}{8z^4} \left(
1+\frac{z^2}{3} - \frac{\tan z}{z}
\right).\label{renR}
\ee
We thus see that to this order the periodic potential produces an overall
$k_x$-dependent renormalization of the dispersion:
\begin{align}
\label{Enqky}
E_n(k_x,k_y) = \left[1+ d^4{\cal R}(d|k_x-K_n|) \right] E_n^0(k_x,k_y).
\end{align}
Eq.~(\ref{Enqky}) holds provided $k_x$ 
is not too close to the boundary of the SBZ. In fact, 
for $k_x\simeq \pi/2d$, $\cal R$ diverges due to the last term in Eq.~(\ref{renR}), 
which signals the breakdown of the perturbative calculation. This is simply due to
the fact that close to the SBZ boundary  the state at $K_n$ is quasi-degenerate with 
the state at $K_{-n+1}$ and they are coupled by the perturbation. Therefore,
one should use degenerate perturbation theory.
It is easy to see that at $k_x=\frac{\pi}{2d}$, $k_y=0$,
the perturbation opens a gap of size $2|A_{-(2n-1)\pi/d}|=4d/\pi^2 (2n-1)^2$,
 which decreases with increasing $n$ (see Fig. \ref{fig.4}).

From Eq.~(\ref{Enqky}) we see that the positions of the finite-energy Dirac points 
coincide with those found at high energies, Eq. (\ref{Bzero}), 
up to a correction of order $d^4$, namely\cite{footnoteEnergies} 
$\frac{|n|\pi}{d}\left[1+d^4{\cal R}(n\pi)\right]$,
where
\be
{\cal R}(\pi n)=\left\{\begin{array}{lc}
-\frac{1}{60}  & \text{for} \;   n = 0, \\
\left( 1+\frac{3}{\pi^2n^2} \right)\frac{1}{24\pi^2n^2} & \text{for} \; n\neq 0.
\end{array}
\right.
\ee
Notice that, due to the smallness of ${\cal R}$ away from the SBZ boundary,
the range of validity of the perturbative calculation actually extends to values 
of $d$ of order $1$.
We can also explicitly compute the velocities at the Dirac points and find 
\begin{align}
v_{nx} &=
\left\{\begin{array}{lc}
1-\frac{d^4}{60}  & \text{for} \;   n = 0, \\
1- \frac{d^4}{24\pi^2n^2}\left(1+\frac{12}{\pi^2 n^2}\right) & \text{for} \; n \neq0,
\end{array}
\right.\\
v_{ny}&=\left\{\begin{array}{lc}
1-\frac{d^4}{60}  & \text{for} \;   n = 0, \\
0 & \text{for} \; n\neq 0.
\end{array}
\right.
\end{align}
In particular, in the case $n=0$ we recover the small-$d$ expansion 
of the exact result, Eq.~(\ref{vsmalld}). Interestingly, within this perturbative 
calculation the $k_y$-dispersion at any $n>0$ and $k_x=0$
remains massive, $E_n(0,k_y) \propto \sqrt{K^2_n+k_y^2}$,
and hence the corresponding velocity $v_{ny}$ always vanishes at $k_y=0$.

\subsection{Case $d\gg 1$}
\label{largeB}

Let us now present the calculation of the spectrum in the 
limit $d\gg 1$, where a tight-binding approximation is justified. 
In fact, in this limit the wavefunctions are well localized deeply in the center of a region of uniform $B$.
In this "atomic" limit the energy eigenvalues are simply the Landau levels
$E_n=\pm \sqrt{2n}$, $n=0,1,2,\dots$, which are (infinitely) doubly degenerate, 
since for each energy there are two eigenstates per superlattice unit cell, corresponding to Landau orbitals in the $B_z>0$ region and in the $B_z<0$ region. 
The degeneracy is lifted for finite $d$ (except at $k_x=k_y=0$, where it is protected by
an exact  symmetry) because the wavefunctions have (exponentially)
small overlaps. We thus expect that the leading correction to the
Landau level $E_n$ originates from the hopping between adjacent degenerate 
Landau orbitals. In order to calculate such correction  
we make the following ansatz for the wavefunction:\cite{ashcroft}
\begin{align}
\psi(x) = &\sum_{R_A,n} e^{ik_x R_A}  a_n \Phi_{n,r}(x -  R_A(k_y)) + \nonumber  \\
 &+ \sum_{R_B,n}  e^{ik_x R_B}  b_n  \Psi_{n,r}(x -  R_B(k_y)),
\end{align}
where $a_n,\, b_n$ are complex coefficients.
$\Phi_{n,r}(x)$ (resp. $\Psi_{n,r}(x)$), with $n=0,1,2,\dots$ and $r=\pm1$,
are the  two-component  relativistic Landau
orbitals in positive (resp. negative) uniform magnetic field 
with energy $E=r\sqrt{2n}$:
\begin{align}
&\Phi_{n,r}(x) 
= C_n
\left(
\begin{array}{c}
\phi_{n-1}(x)\\
r \phi_n(x)
\end{array}
\right), \\
&\Psi_{n,r}(x) =\sigma_x \Phi_{n,r}(x), \\
&C_0=1, \quad C_{n>0}=\frac{1}{\sqrt{2}}.
\end{align}
The functions $\phi_n$ are the harmonic oscillator eigenstates
\be
\phi_n(x) = \sqrt{\frac{1}{2^n n!}}\left( 
\frac{1}{\pi} \right)^{1/4} H_n(x)\, e^{-x^2/2}
\ee
with $H_n(z)$ the Hermite polynomials.
(For $n=0$ it is understood that $\phi_{-1}\equiv 0$ and there is no index $r$.)
Finally, $R_A(k_y)\equiv R_A -  k_y= 2jd +\frac{d}{2} - k_y$ 
and $R_B(k_y)\equiv R_B +  k_y= (2j+1)d +\frac{d}{2} + k_y$  ($j\in \mathbb{Z}$)
denote the shifted centers of the LL orbitals.

We now keep into account only the hopping between nearest-neighbor orbitals
and focus on the level $E_n$. In the two-dimensional subspace of degenerate 
levels, under usual approximations, the DW equation reduces to
\begin{align}
\left( \begin{array}{cc}
E_n-E  & \Delta_n(k_x,k_y) \\
\Delta_n^*(k_x,k_y) & E_n -E
\end{array}
\right) 
\left( \begin{array}{c}
a_n\\
b_n
\end{array}
\right) = 0,
\end{align}
where $\Delta_n(k_x,k_y)$ is the hopping matrix element
\begin{widetext}
\begin{align}
\Delta_n(k_x,k_y)  = \sum_{R_B=R_A \pm 1} e^{-ik_x (R_A-R_B)} \int dx \, 
\Phi_{n,+} (x-R_A(k_y)) H \Psi_{n,+} (x-R_B(k_y)),\label{matrixel}
\end{align}
\end{widetext}
with the DW Hamiltonian $H$ given in Eq.~(\ref{ham1d}).
The matrix element (\ref{matrixel})
can be evaluated by using the fact that the dominant
contribution to the integral originates from the region around the interface
between domains of  opposite magnetic field.
We then find the energy eigenvalues
\begin{align}
E &=E_n\pm |\Delta_n(k_x,k_y)|  \label{En_large_d} \\
& = \sqrt{2n} \pm C_n^2\left[
{\cal A}_n^2(k_y) \cos^2 (k_xd)  + {\cal B}_n^2(k_y)\sin^2 (k_x d)
\right]^{1/2}
\nonumber 
\end{align}
where
\begin{align}
{\cal A}_n(k_y) &= \sum_{r=\pm 1} r \left[ \phi_n^2(k_y +rd/2) - \phi_{n-1}^2(k_y +rd/2)\right]
\label{An}\\ 
{\cal B}_n(k_y) & =\sum_{r=\pm 1} \left[ \phi_n^2(k_y +rd/2) - \phi_{n-1}^2(k_y +rd/2)\right].
\label{Bn}
\end{align}
Equation~(\ref{En_large_d}) holds throughout the superlattice Brillouin zone and
for any $k_y$ provided $k_y/d \ll 1/2$. The last condition ensures that the shifted 
centers of the Landau orbitals are far from the interfaces, which justifies some of 
the approximations used in the calculation. For $k_y \simeq d/2$ the states 
transmute into snake states, which we do not discuss further in this paper. 

At $k_x=k_y=0$ $\Delta_n$ vanishes, as it should be, since the degeneracy
at this point is protected by symmetry. Expading for small $k_x$ and $k_y$
we recover a Dirac conical  dispersion centered at $E_n=\sqrt{2n}$ as 
in Eq.~(\ref{dispersion}), with velocities given by
\begin{align}
v_{nx} &=C^2_n {\cal B}_n(0)\,d , \\
v_{ny} &= C^2_n \left| {\cal A}'_n(0) \right|,
\end{align}
whose explicit expressions can be obtained from Eqs.~(\ref{An}) and (\ref{Bn})
and nicely agree with the results of Sec. \ref{larged}, Eqs.~(\ref{v_nx_asy}) and 
(\ref{v_ny_asy}).


\section{Conclusions}
\label{sec7}

We have shown that in graphene an alternating magnetic field, 
whose modulation has a typical length scale much larger than the lattice constant,
does not spoil the Dirac cone dispersion close to zero energy and, moreover, 
induces new Dirac points in the spectrum at higher energies. 
The positions of the new singular points scale at first as $\sqrt{n}$, in analogy 
to relativistic Landau levels, but for larger energies cross over to a linear 
dependence on $n$.
Surprisingly, despite the strong anisotropy of the field profile, the quasi-particle dispersion 
around zero energy is still isotropic. On the contrary, at the higher Dirac points, 
the group velocity components in the directions parallel and perpendicular to the 
superlattice strongly differ. There exists a parameter regime where the dispersion 
along the interfaces is substantially suppressed, as shown in Fig.~\ref{fig.5}. 
As a result, close to these new points, the DW quasi-particles propagate, rather
counterintuitively, more easily in the direction perpendicular to the magnetic barriers 
than along them, for $v_{n x}$ is always greater than $v_{n y}$.  
One may therefore exploit this effect to focus and collimate a quasi-particle beam by suitably tuning the doping
in such a way that the Fermi level reaches one of these anisotropic Dirac points.
 
The robustness of these new Dirac points in the presence of various types of 
disorder and the implications of their existence on the transport properties
of graphene's magnetic superlattice are interesting topics, that we hope to 
address in the near future.


\acknowledgments

We thank R. Egger for a critical reading of the manuscript. 
A.D.M. acknowledges the financial support of the SFB/TR 12 of the DFG.



\begin{thebibliography}{99}

\bibitem{reviews} 
For recent reviews, see 
A.K. Geim and K.S. Novoselov, Nature Materials {\bf 6}, 183 (2007); 
A. Geim, Science {\bf 324}, 1530 (2009);
A.H. Castro Neto, F. Guinea, N.M.R. Peres, K.S. Novoselov, and A.K. Geim, 
Rev. Mod. Phys. {\bf 81},109 (2009).

\bibitem{park2008b}
C.-H. Park, L. Yang, Y.-W. Son, M.L. Cohen, and S.G. Louie,
Nat. Phys. {\bf 4}, 213 (2008). 

\bibitem{park2008c}
C.-H. Park, L. Yang, Y.-W. Son, M.L. Cohen, and S.G. Louie,
Nato Lett. {\bf 8}, 2920 (2008). 

\bibitem{park2008a}
C.-H. Park, L. Yang, Y.-W. Son, M.L. Cohen, and S.G. Louie,
Phys. Rev. Lett. {\bf 101}, 126804 (2008). 

\bibitem{park2009}
C.-H. Park, Y.-W. Son, L. Yang, M.L. Cohen, and S.G. Louie,
Phys. Rev. Lett. {\bf 103}, 046808 (2009). 

\bibitem{brey2009}
L. Brey and H.A. Fertig, 
Phys. Rev. Lett. {\bf 103}, 046809 (2009).

\bibitem{barbier2010}
M. Barbier, P. Vasilopoulos, and F.M. Peeters,
Phys. Rev. B {\bf 81}, 075438  (2010).

\bibitem{nogaret2010}
For a recent review 
see A. Nogaret,
J. Phys.: Condens. Matter {\bf 22}, 253201 (2010). 

\bibitem{bao2009}
W. Bao {\it et al.}, Nature Nanotechnology {\bf 4}, 562 (2009).

\bibitem{guinea2008}
F. Guinea, M.I. Katsnelson, and M.A.H Vozmediano,
Phys. Rev. B {\bf 77}, 075422 (2008).

\bibitem{pereira2009}
V.M. Pereira and A.H. Castro Neto,
Phys. Rev. Lett. {\bf 103}, 046801 (2009).

 \bibitem{luca2009}
 L. Dell'Anna and A. De Martino,
 Phys. Rev. B {\bf 79}, 045420 (2009); {\bf 80} 089901(E) (2009). 

\bibitem{snyman2009}
I. Snyman, 
Phys. Rev. B {\bf 80}, 054303 (2009).

\bibitem{louie2010} 
L.Z. Tan, C.-H. Park, and S.G. Louie,
Phys. Rev. B {\bf 81}, 195426 (2010).

 \bibitem{ghosh2009}
 S. Ghosh and M. Sharma, 
 J. Phys.: Condens. Matter {\bf 21}, 292204 (2009).

 \bibitem{ramezani2009}
M. Ramezani Masir, P. Vasilopoulos, and F.M. Peeters,
New Jour. Phys. {\bf 11}, 095009 (2009).
 
 \bibitem{ramezani2010}
 M.Ramezani Masir, P. Vasilopoulos, and F.M. Peeters,
 J. Phys. Condens. Matter, {\bf 22} 465302 (2010).

\bibitem{xu2010a}
L. Xu, J. An, and C.-D. Gong,
Phys. Rev. B {\bf 81}, 125424 (2010).

\bibitem{xu2010b} 
L. Xu, J. An, and C.-D. Gong,
Phys. Rev. B {\bf 82}, 155421 (2010).

\bibitem{belzig2010}
S. Gattenl\"ohner, W. Belzig, and M. Titov,
Phys. Rev. B {\bf 82}, 155417 (2010).

\bibitem{ale} A. De Martino, L. Dell'Anna, and R. Egger, 
Phy. Rev. Lett. {\bf 98}, 066802 (2007); 
Sol. State Comm. {\bf 144}, 547 (2007).

\bibitem{grad} 
I.S. Gradshteyn and I.M. Ryzhik, {\sl Table of Integrals, Series, and Product}
(Academic Press, Inc., New York, 1980).

\bibitem{tarun} 
T.K. Ghosh, A. De Martino, W. H\"ausler, 
L. Dell'Anna, and R. Egger,
Phys. Rev. B {\bf 77}, 081404(R) (2008).

\bibitem{cserti} 
P. Rakyta, L. Oroszlany, A. Kormanyos, 
C.J. Lambert, and J. Cserti, 
Phys. Rev. {\bf 77}, 081403(R) (2008). 

\bibitem{footnote} Equivalently, one can identify 
the perturbative parameter as the (square root) of 
the phase acquired by a quasiparticle moving around a 
plaquette of side $d$, $2\pi Bd^2/\Phi_0$, with $\Phi_0=hc/e$ the flux quantum.

\bibitem{watson} 
G.N. Watson, 
Proc. London Math. Soc. (2) {\bf 8}, 393 (1910); see also
N. Schwid, 
Trans. Amer. Math. Soc. {\bf 37}, 339 (1935).

\bibitem{lucaspin} 
L. Dell'Anna and A. De Martino,
Phys. Rev. B {\bf 80}, 155416 (2009).

\bibitem{ashcroft} 
See, e.g., N.W. Ashcroft and N.D. Mermin,
{\sl Solid State Physics}
(Thomson Learning, Inc., 1976).

\bibitem{footnoteEnergies} Remember that we focus on the non-negative
part of the spectrum. 

\end{thebibliography}
\end{document}